\documentclass[aps,prl,twocolumn,superscriptaddress,groupedaddress]{revtex4}  

\usepackage{graphicx,color}  
\usepackage{dcolumn}   
\usepackage{bm}        
\usepackage{amssymb}   

\usepackage[section]{placeins}
\usepackage{float}
\usepackage{epsfig,amssymb,amsmath,amsthm,amsfonts,amsbsy,mathrsfs}
\usepackage{multirow}
\usepackage{array}

\begin{document}

\title{Optimal segregation of proteins: phase transitions and symmetry breaking}
\author{Jie Lin}
\affiliation{School of Engineering and Applied Sciences, Harvard University, Cambridge, Massachusetts 02138, USA}

\author{Jiseon Min}
\affiliation{School of Engineering and Applied Sciences, Harvard University, Cambridge, Massachusetts 02138, USA}
\affiliation{Department of Physics, California Institute of Technology, Pasadena, California 91125, USA}

\author{Ariel Amir}
\affiliation{School of Engineering and Applied Sciences, Harvard University, Cambridge, Massachusetts 02138, USA}

\date{\today}
\begin{abstract}
Asymmetric segregation of key proteins at cell division -- be it a beneficial or deleterious protein -- is ubiquitous in unicellular organisms and often considered as an evolved trait to increase fitness in a stressed environment. Here, we provide a general framework to describe the evolutionary origin of this asymmetric segregation.
We compute the population fitness as a function of the protein segregation asymmetry $a$, and show that the value of $a$ which optimizes the population growth manifests a phase transition between symmetric and asymmetric partitioning phases. Surprisingly, the nature of phase transition is different for the case of beneficial proteins as opposed to proteins which decrease the single-cell growth rate. Our study elucidates the optimization problem faced by evolution in the context of protein segregation, and motivates further investigation of asymmetric protein segregation in biological systems.
\end{abstract}
\maketitle

{\it Introduction - }In stressed environments, microbial cells such as bacteria or yeast utilize various mechanisms in order to survive. One important mechanism is the asymmetric segregation of vital cytosolic components at cell division: one of the two daughter cells will inherit more favorable conditions (at the expense of the sister cell) \cite{Stewart2005,Lindner2008,Vedel2016, Coelho2014,Vaubourgeix2015,Saarikangas2017}. Experiments on fission yeast have shown that this asymmetric segregation emerges in a stressed environment but not under favorable conditions \cite{Coelho2013}. These observations led to a conjecture that cells may have evolved asymmetric segregation of deleterious damages to increase the overall fitness of the population since one of the daughter cells would be ``rejuvenated" \cite{Watve2006,Ackermann2007,Chao2016,Vedel2016}. Cytosolic components that are advantageous to cell growth may also be segregated asymmetrically. For instance, recent experiments elucidated the molecular mechanisms of asymmetric segregation of the main multidrug efflux pumps in the bacterium {\it Escherichia coli}, which enable direct expulsion of harmful chemicals from the cells \cite{Bergmiller2017}. The strong partitioning bias of efflux pumps for the old cell poles generates growth rate differences among cells, and it has been argued to be a strategy for bacteria to survive in the presence of a high concentration of antibiotics.

Several previous theoretical works focused on particular models for how damage affects cellular growth, in which it was shown that completely asymmetric segregation (one daughter cell inherits all the key protein from the mother cell while the other becomes free of the protein) optimizes the population growth rate \cite{Chao2010, Vedel2016,Ackermann2007}. Here, we take a different approach to the problem -- rather than focusing on a particular model, we will study a rather broad class of models in which the instantaneous (single-cell) growth rate is a function of the protein concentration. Notably, we will find the optimal segregation strategy that maximizes the population growth rate. In this way, we obtain general insights into the problem and identify the underlying optimization principles. Also, previous studies often consider a coarse-grained effect of the damaged proteins on the cell's fitness through the generation time \cite{Chao2010,Vedel2016} or the survival rate \cite{Ackermann2007}, neglecting the exponential growth of cell volume at the single level \cite{Wang2010,Campos2014,Taheri2015,Cermak2016}. For many scenarios a more realistic model should start from the instantaneous effects of the key protein cellular concentration on the single-cell growth rate.




\begin{figure}[htb!]
	\includegraphics[width=0.48\textwidth]{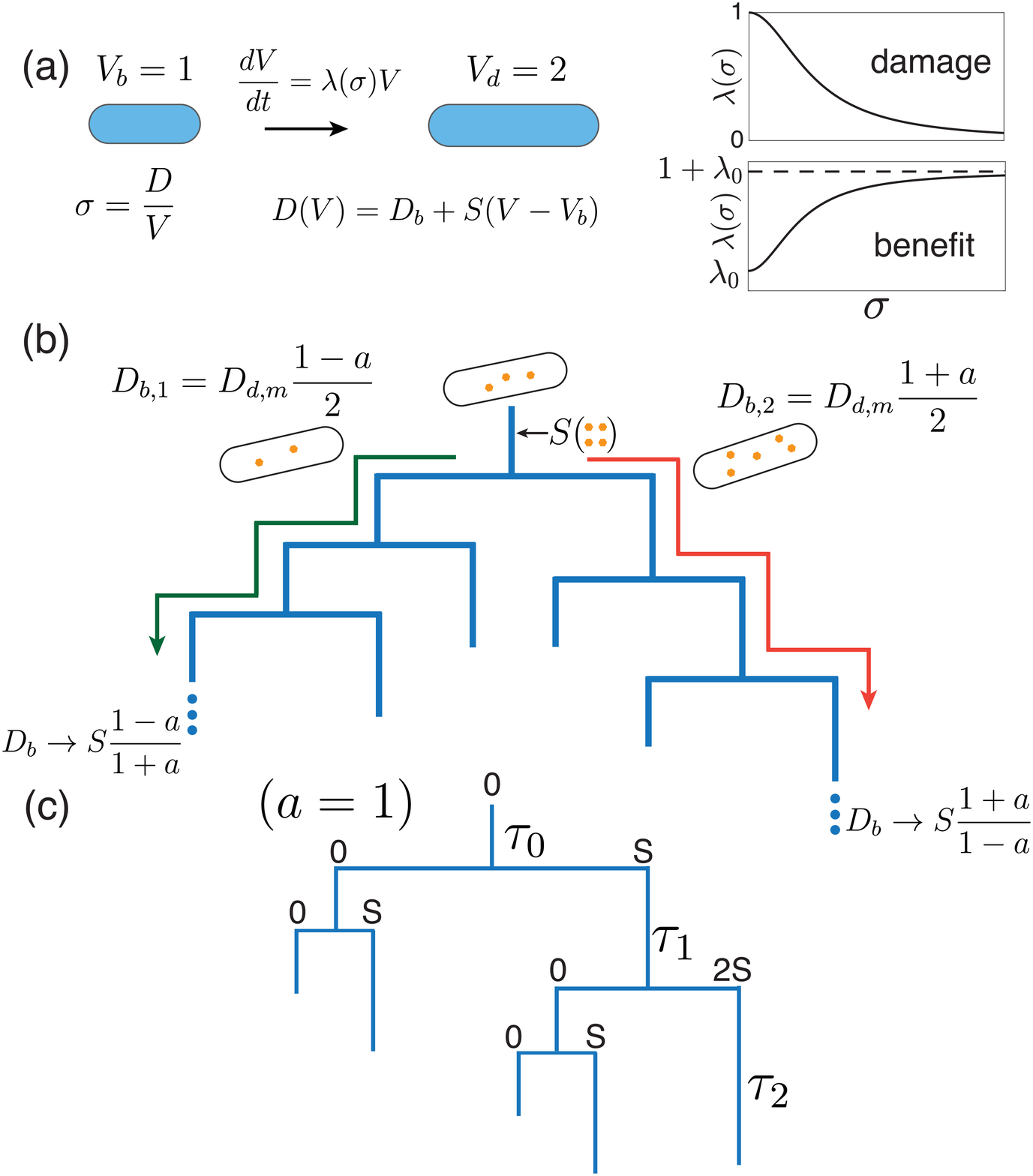}
	\caption{(a) (Left) Each cell is born with the volume $V_b=1$ and divides with the volume $V_d=2$. The amount of newly produced key protein is proportional to the increment of cell volume with the proportionality  constant $S$. The key protein concentration is $\sigma=D/V$. (Right) The instantaneous growth rate depends on the key protein concentration, which can be deleterious (damage) or beneficial (benefit). (b) A mother cell attributes the protein asymmetrically to the daughters, with an asymmetry parameter $a$. The lineage composed of cells inheriting more proteins than its sister cell will reach a steady state with a constant number of proteins at cell birth (red arrow). Similarly, the other lineage composed of cells inheriting less proteins (green arrow) will reach a steady state. (c) The amounts of key protein at cell birth are shown for a population tree with $a=1$. In this case, the whole tree can be decomposed into subtrees, which are identical to the entire tree but shifted by a delay time, $\tau_0$, $\tau_0+\tau_1$, etc.}\label{figure1}
\end{figure}

For the completely asymmetric segregation case, we derive an analytical expression for the population growth rate. For weak asymmetry, we can map the problem to the Landau theory of phase transitions \cite{Chaikin2000}. Interpolating these two regimes, we are able to find the optimal segregation ratio which maximizes the population growth, for a given level of environmental stress. We find that the optimal ratio exhibits a phase transition from a symmetric phase to a perfectly asymmetric phase as the environmental stress increases. While the transition is sharp for the case of deleterious proteins, for the segregation of benefits the transition is of second order. These theoretical predictions are verified using numerical simulations. We conclude by discussing the relation between our theory and experimental observations.

{\it Model-} We summarize the model in Fig. \ref{figure1}(a,b). We assume that the amount of the newly produced key protein is proportional to the increment of cell volume, with an accumulation rate $S$. Therefore, the amount of protein at cell volume $V$ is equal to
\begin{equation}
D(V)=D_b+S(V-V_b),\label{Dv}
\end{equation}
and here $V_b$ ($D_b$) is the cell volume (amount of protein) at cell birth. The protein accumulation rate $S$ quantifies the environmental stress: a larger $S$ represents a more stressed environment for damage segregation, {\it e.g.}, a higher temperature \cite{Coelho2013} and a less stressed environment for benefit segregation. In the case of beneficial proteins, the usual parameters tuned in experiments is the concentration of antibiotics \cite{Bergmiller2017}. A higher concentration of antibiotics requires a higher concentration of beneficial proteins, {\it e.g.}, the efflux pump, to achieve the same growth rate as a lower concentration of antibiotics. So increasing the environmental stress through antibiotics concentration is equivalent to lowering the beneficial protein concentrations, which is set by the protein accumulation rate. In the following, we consider the protein accumulation rate as the control parameter for both damage and benefit segregation cases, and a larger (smaller) $S$ represents a more (less) stressed environment for damage (benefit) segregation.

We assume that the cell divides its volume symmetrically and deterministically ($V_b = 1, V_d = 2$) based on the fact that the cell volume fluctuation is small \cite{Amir2014,Campos2014,Taheri2015,Ho2018}. Each dividing cell allocates the key protein to the two daughter cells, and the amounts of protein that the two daughter cells inherit are
\begin{equation}
D_{b,1}=D_{d,m}\frac{1-a}{2},\quad
D_{b,2}=D_{d,m}\frac{1+a}{2},\label{Db}
\end{equation}
where $D_{d,m}$ is the amount of protein inside the mother cell at division, and $a$ is a continuous variable, ranging from $-1$ to $1$ (Fig. \ref{figure1}(b)). Because the two daughter cells are equivalent in our model, we choose $ 0\leq a \leq 1$. $a=1$ corresponds to the completely asymmetric segregation and $a=0$ corresponds to the symmetric segregation.



We assume that the cell volume grows exponentially at the single cell level \cite{Wang2010, Campos2014,Cermak2016},	
\begin{equation}
\frac{dV}{dt}=\lambda[\sigma]V,\label{dV}
\end{equation}
where $\lambda[\sigma]$ is the instantaneous growth rate, depending on the protein concentration $\sigma=D/V$.
The growth rate function depends on the specific system, which can be measured experimentally and in general exhibits an inflection point \cite{Lindner2008,Vedel2016}. Here, we consider the growth rate as a Hill-type function,
\begin{equation}
\lambda[\sigma]=\frac{\lambda_0 + \lambda_1 \sigma^n}{1+\sigma^n}.\label{lambda}
\end{equation}
If $\lambda_0 > \lambda_1$, $\lambda[\sigma]$ is a decreasing function, and thus the key protein is deleterious. If $\lambda_0 < \lambda_1$, the protein is beneficial. In this paper, we focus on the case $n=2$, but our main conclusions are valid for any $n > 1$ (Supplementary Information (SI) (b)).

First, we discuss the special case $n=1$, where the growth rate function is purely convex (concave) for the damage (benefit) case. Assume that a mother cell has a protein concentration $\sigma$ and compute the difference in the growth rate of total cell volume right before and after the division: $\Delta=\lambda[\sigma(1-a)]+\lambda[\sigma(1+a)]-2\lambda[\sigma]$. If the growth rate function is convex (concave), setting $a=1$ ($a=0$) always maximizes $\Delta$ for every division event. Therefore it is plausible that the population growth rate is also always maximized at $a=1$ ($a=0$) independent of the environmental stress (see numerical tests in SI (a)). This is consistent with the previous work on related models \cite{Ackermann2007,Vedel2016}.

For a general growth rate function with an inflection point, however, the optimal $a$ for each single division event depends on the mother cell's damage concentration $\sigma$. In other words, considering a single division event is not sufficient for finding the optimizing $a$. In the following, we provide two methods to find the population growth rate respectively for small $a$ and $a=1$. Interpolating between these two limits will provide insights into the optimal degree of asymmetry.
\begin{figure}[htb!]
   \includegraphics[width=.47\textwidth]{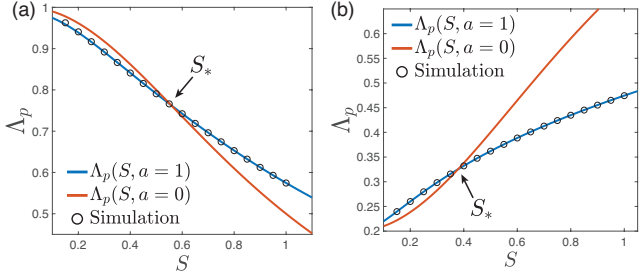}
    \caption{The population growth rate at $a=0$ and $a=1$ for the damage (a) and benefit (b) case. The blue solid lines are theoretical results based on numerical calculation of Eq. (\ref{recursion1}), and the red solid lines are $\lambda[S]$. The circles are results from direct simulations of a population with $a=1$. $S_{\ast}$ is the accumulation rate at which $\Lambda_p(S_{\ast},0)=\Lambda_p(S_{\ast},1)$. The growth rate at the single-cell level is equal to $\lambda[\sigma]=\frac{\lambda_0 + \lambda_1 \sigma^2}{1+\sigma^2}$ with $\lambda_0=1$, $\lambda_1=0$ for the damage case and $\lambda_0=0.2$, $\lambda_1=1.2$ for the benefit case. Note that $\lambda_0>0$ for the benefit case in order to ensure a well defined population growth rate also for perfectly asymmetric partitioning. The value of $\lambda_0$ does not affect our conclusions (see SI (d)). $S_{\ast}\approx0.558$ for the damage case and $S_{\ast}=0.378$ for the benefit case.
    }\label{figure2}
\end{figure}

{\it Self-similarity method - }We consider an exponentially growing population and imagine taking a snapshot of the population at some time, from which we can find the total number of cells ($N$) and the total cell volume of the whole population ($V_t$). The population growth rate must be equal to the total volume growth rate $\frac{1}{N}\frac{dN}{dt}=\frac{1}{V_t}\frac{dV_t}{dt}=\Lambda_p$ because the cell volume is regulated. We use the total cell volume to compute $\Lambda_p$ because it is more accurate and tractable to measure the growth rate of the total cell volume than total cell number in numerical simulations and analytical analysis.

When $a=0$, the concentration of key protein remains constant as the cell is growing from $V=1$ to $V=2$, and is equal to the accumulation rate of key protein $S$. Therefore, all cells grow at the same rate $\lambda[S]$, and the population growth rate is equal to the homogeneous single-cell growth rate $\Lambda_p=\lambda[S]$.

When $a=1$, one of the daughter cells does not get any of the key protein from its mother, while the other one inherits all the key protein. This leads to a self-similarity in the population tree (Fig. \ref{figure1}(c)). Consider a tree starting from a single cell without any key protein. The whole tree can then be decomposed into subtrees, which are identical to the original one, except for a temporal shift. The number of cells of the whole tree grows as $N(t)\sim \exp(\Lambda_p t)$. For the subtrees, they share the same $\Lambda_p$ as the whole tree, but with a temporal shift, $\sum_{j=0}^{i} \tau_j(S))$ ($i=0,1,2,...$), where $\tau_j(S)$ is the generation time of a cell whose amount of key protein at cell birth is $jS$ (SI (c)). The self-consistent equation of the population growth rate is
\begin{equation}
\sum_{i=0}^{\infty} \exp\Big(-\Lambda_p\sum_{j=0}^{i}\tau_j\Big) =1 . \label{recursion1}
\end{equation}
The above equations are satisfied by a unique value of population growth rate $\Lambda_p$. In Fig. \ref{figure2}(a,b), we plot the numerical values of $\Lambda_p$ when $a=1$, based on Eq. \ref{recursion1}.
We note that both for the damage and benefit cases, there exists a special $S_{\ast}$ at which $\Lambda_p(S_{\ast}, a = 1) = \Lambda_p(S_{\ast}, 0)=\lambda[S_{\ast}]$.  For the damage (benefit) case, $a = 0$ is favored to $a = 1$ if $S < S_{\ast}$ ($S > S_{\ast}$), and $a = 1$ is favored to $a = 0$ if $S > S_{\ast}$ ($S < S_{\ast}$) (Fig. \ref{figure2}(a,b)). Yet, comparing the two exactly solvable cases is not sufficient for finding the optimal $a_c$ which maximizes the population growth rate. Moreover, this comparison does not show how $a_c$ changes with the control parameter $S$. In the following section, we will introduce the Landau approach and show that $S_{\ast}$, with the inflection point $S_c$ of the growth rate function, determines the nature of the transition from $a_c=0$ to $a_c=1$.

\begin{figure}[htb!]
   \includegraphics[width=.45\textwidth]{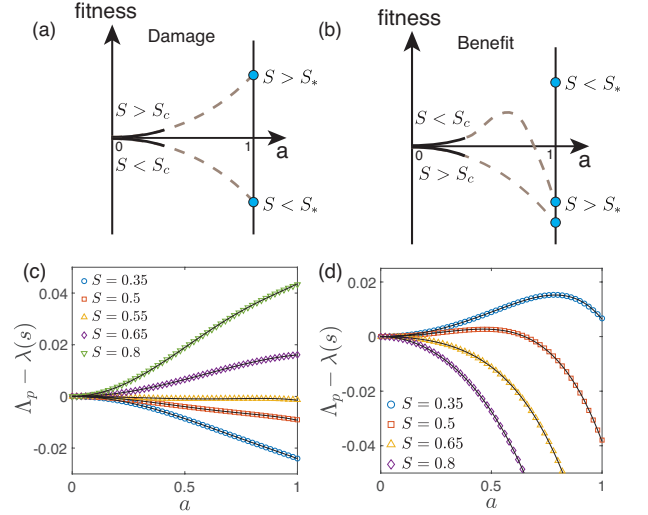}
    \caption{(a) For the damage case, because $S_{\ast}\approx S_c$, the fitness at $a = 1$ and its curvature at $a = 0$ flip their signs simultaneously. This leads to a sharp transition of the optimal asymmetry parameter $a_c$ from $1$ to $0$. (b) For the benefit case, because $S_{\ast}$ is far below $S_c$, as one increases $S$ from a low value, a small finite $a$ must exist that maximizes the fitness if $S$ is just below $S_c$  This leads to a smooth second order transition with a mean field exponent, $a_c\sim(S_c-S)^{1/2}$. (c,d) Numerical simulations of the fitness as function of $a$ for the damage (c) and benefit (d) cases. The black lines are the fits based on $f=C_2a^2+C_4a^4+C_6a^6$. }\label{figure3}
\end{figure}

{\it Landau approach - }We next consider the general case with $a\neq 0$. We decompose the growth of total cell volume ($V_t$) as the sum of all individual cell contributions, $dV_t/dt=\sum_i\lambda[\sigma_i]V_i$. Because when $a=0$, all cells have the same protein concentration $S$, we choose to expand around $\sigma_i = S$ for each cell,
\begin{align}
\frac{dV}{dt}
=\sum_i \lambda[\sigma_i]V_i
= V \lambda[S]+ \sum_i \frac{d\lambda}{d\sigma}\Bigr |_S(\sigma_i-S)V_i\nonumber \\
+\frac{1}{2}\frac{d^2\lambda}{d\sigma^2}\Bigr |_S(\sigma_i-S)^2V_i+\frac{1}{3!}\frac{d^3\lambda}{d\sigma^3}\Bigr |_S(\sigma_i-S)^3V_i+...\nonumber
\end{align}
The first order term vanishes because the total amount of protein of the entire population, $\sum \sigma_i V_i=SV_t$ in the steady state. We define the population's fitness as $f=\Lambda_p-\lambda[S]$,
\begin{align}
f(S,a)&=\frac{1}{2}\frac{d^2\lambda}{d\sigma^2}\Bigr |_S\langle (\sigma_i-S)^2\rangle_v+\frac{1}{3!}\frac{d^3\lambda}{d\sigma^3}\Bigr |_S\langle (\sigma_i-S)^3\rangle_v+...\nonumber
\end{align}
where $\langle (\sigma_i-S)^n\rangle_v\equiv (\sum_i (\sigma_i-S)^nV_i)/\sum_i V_i$. Note that the fitness is directly related to the population's phenotypic heterogeneity -- the fluctuation of protein concentration. In the symmetric case ($a=0$), the fitness $f$ is zero by definition. Consider now the limit $0<a\ll1$. Because in the limit of small $a$, $\langle (\sigma_i-S)^n\rangle_v\sim a^n$, the lowest order term of the fitness function has to scale as $a^2$ with its coefficient proportional to the second derivative of the growth rate function at $S$,
\begin{equation}
f=A S^2\frac{d^2\lambda}{d\sigma^2}\Bigr |_S a^2+C_4a^4+C_6 a^6+O(a^8).\label{landau}
\end{equation}
Here $A$ is a universal positive number independent of the growth rate function (SI (f)). The odd order terms vanish because of the symmetry $f(a)=f(-a)$, and $C_4$, $C_6$ are constants depending on the second and higher order derivatives of the growth rate function. We therefore see that the sign of the second derivative of the growth rate function determines whether the symmetric phase $a=0$ is a local maximum or minimum of the fitness. The analysis which follows is reminiscent of the Landau approach to ferromagnetic phase transitions: the free energy is replaced by the fitness and the magnetization is replaced by the asymmetry parameter. The inflection point $S_c$ of the growth rate function, at which its second derivative vanishes, will also be central to the analysis.


For the damage case, we find $S_{\ast}\approx S_{c}$ ($S_{\ast}=0.558$, $S_c=0.577$, Fig. \ref{figure2}(a)) taking $\lambda_0=1$, $\lambda_1=0$. We numerically confirm this is true for other Hill exponents $n$ (SI (b)) and prove $S_{\ast}\approx S_{c}$ if $\lambda_1=0$ (SI (c)). As one increases the protein accumulation rate, the fitness at $a=1$ changes from negative to positive when $S$ exceeds $S_{\ast}$, and the curvature of the fitness at $a=0$ changes from negative to positive when $S$ exceeds $S_c$ (Fig. \ref{figure3}(a)). Therefore, if $S_{\ast}<S_c$, as one increases $S$, the fitness at $a=1$ changes from negative to positive before the curvature at $a=0$ changes its sign. As a result, $a_c$ should undergo a sharp transition from $0$ to $1$ as $S$ increases. In particular, the transition of $a_c$ should be sharp as well if $S_{\ast}\approx S_c$ (Fig. \ref{figure2}(c)), because the fitness at $a=1$ and its curvature at $a=0$ flip their signs simultaneously (Fig. \ref{figure3}(a)). In SI (d), we also numerically confirm $S_{\ast}<S_c$ for a finite $\lambda_1$, therefore a sharp transition is generally true for a Hill function which decreases monotonically. 

For the benefit case, we find $S_{\ast}<S_c$ ($S_{\ast}=0.378$, $S_c=0.577$, Fig. \ref{figure2}(b)) with $\lambda_0=0.2$, $\lambda_1=1.2$. We numerically confirm that $S_{\ast}<S_c$ for other choices of Hill exponent (SI (b)) and $\lambda_0$ (SI (c)). We also prove $S_{\ast}<S_c$ in the limit $\lambda_0\rightarrow 0$ (SI (d)). As one increases $S$ from a low value, the fitness at $a=1$ changes from positive to negative when $S$ exceeds $S_{\ast}$ and the curvature at $a=0$ changes from positive to negative when $S$ exceeds $S_c$ (Fig. \ref{figure3}(b)). Therefore, if $S_{\ast}<S_c$, a small finite $a$ must exist that maximizes the fitness if $S$ is just below $S_c$, and thus we predict a smooth transition of the optimal $a_c$ (Fig. \ref{figure3}(b)). Moreover, by finding the optimal $a_c$ that maximize the fitness using Eq. (\ref{landau}), we predict that the smooth transition of the benefit case is in the universality of the Landau mean field model, namely, $a_c\sim |S-S_c|^{1/2}$ \cite{Chaikin2000}. Our conclusion regarding the sharp transition of $a_c$ for the damage case and the smooth transition for the benefit case is is generally true for different types of growth rate function (SI (g)).


\begin{figure}[htb!]
   \includegraphics[width=.45\textwidth]{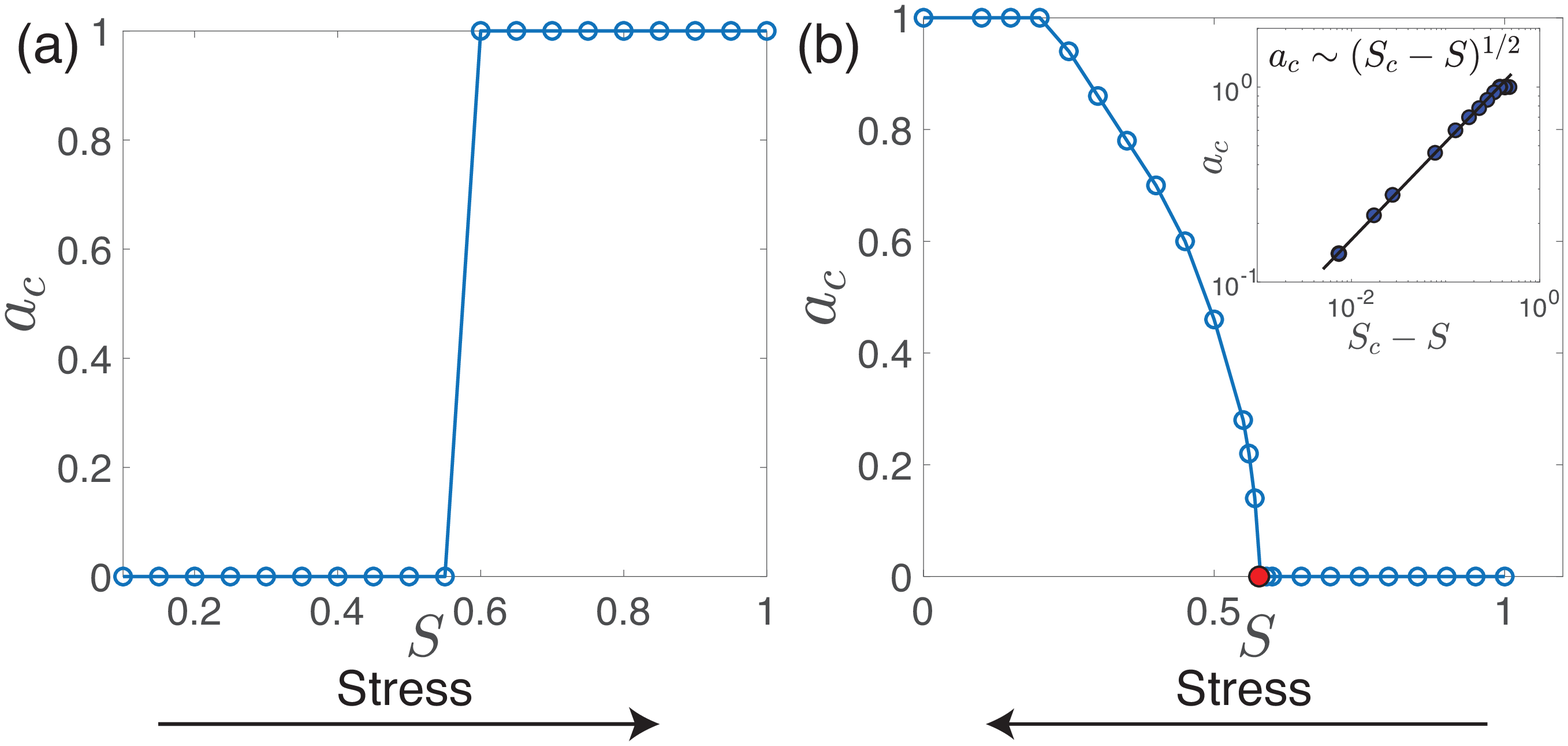}
    \caption{Numerical simulations of the population growth. (a) The optimal $a_c$ changes sharply from $0$ to $1$ at $S_{\ast}$ for the damage segregation. (b) For benefit segregation, the optimal $a_c$ changes smoothly from $1$ to $0$, vanishing at the critical accumulation rate $S_c$, the inflection point of the growth rate function (marked as the red circle). The inset shows the mean field scaling near the critical point where the black line has a slope $1/2$.
    }\label{figure4}
\end{figure}

{\it Numerical simulations - } We test our predictions by simulating an exponentially growing population based on Eqs. (\ref{Dv}-\ref{lambda}) and by setting $(n, \lambda_0, \lambda_1) = (2, 1, 0)$ for the damage case and $(n, \lambda_0, \lambda_1) = (2, 0.2, 1.2)$ for
the benefit case. The simulations start from $100$ cells with random cell volume uniformly distributed from $1$ to $2$ with the same protein concentration $S$. We ran the simulations until there are $5\times 10^{6}$ cells in the population and computed the population growth rate using data after time $t=10$. We compute the fitness as $f=\Lambda_p-\lambda[S]$ and plot it against $a$ in Fig. \ref{figure3}(c,d), which are consistent with Fig. \ref{figure3}(a,b). The optimal $a_c$ that maximizes the fitness changes from $a_c=0$ to $a_c=1$ abruptly for the damage case and smoothly for the benefit case, agreeing with our predictions (Fig. \ref{figure4}(a,b)). For the benefit case, the transition shows a second order mean field behavior as we predict (see inset of Fig. \ref{figure4}(b)). Furthermore, we test our Landau approach by fitting the fitness using $f=C_2a^2+C_4a^4+C_6a^6$ (black lines in Fig. \ref{figure3}(c,d)) and in both of the damage and benefit cases, we obtain the coefficient $A\approx 0. 361$ (SI (e)). In the SI (f), we show that the value of $A$ is universal and determined by the structure of the trees in the limit of small $a$, independent of the specific growth rate function.

{\it  Discussion - }In this paper we studied the optimal segregation strategy of a protein whose presence may be beneficial or deleterious to cellular growth, via combination of analytic and numerical approaches. We found that the optimal degree of asymmetry which maximizes the population growth rate shows a rich behavior and in particular a phase transition taking a different form in the case of damage or benefit segregation: a sharp transition from purely symmetric to asymmetric segregation is found in the former, while a smooth (second order) transition from asymmetric to symmetric segregation occurs in the latter.
Our results are consistent with the segregation of damaged proteins in certain organisms, such as fission yeast in which a transition from symmetric segregation to asymmetric phase is observed as the environmental stress increases \cite{Liu2010, Spokoini2012, Zhou2014, Coelho2013, Coelho2015}. In these organisms, complex molecular machinery, {\it e.g.}, Hsp16, has evolved that actively fuse damaged proteins to achieve completely asymmetrical segregation between the two daughter cells ($a=1$). This suggests that cells may have some control over the segregation ratio by tuning the activity levels of the machinery involved in the asymmetric segregation of damaged protein, making the question we studied here relevant to understanding the optimization problem faced by evolution. Similarly, the beneficial drug pumps, AcrAB-TolC complexes in {\it E. coli} are segregated asymmetrically between the old-pole and new-pole daughter cells in a TolC-dependent manner \cite{Bergmiller2017}. Consistent with our predictions, a continuous change of the asymmetry degree $a$ from $0.06$ to $0.2$ was observed as the subinhibitory antibiotic concentration increases in the experiments.





Our model can be modified to explain asymmetry in various contexts of cell biology. For example, it would be interesting to understand how the asymmetric cell volume division affects the optimal strategy of protein segregation \cite{Soifer2016}. It would also be intriguing to explore how the primarily deterministic effects studied here would be affected by stochasticity \cite{Lin2017}. Furthermore, in some organisms, it might be more accurate to consider $a$ as a function of the current state {\it e.g.} protein concentration $\sigma$ rather than as a free parameter \cite{Coelho2014,Vedel2016}. It would be interesting to investigate the optimal segregation strategy in such a case, where the segregation process is passively controlled by the current protein concentration.

{\bf Acknowledgments}: We thank Yohai Bar-Sinai, Miguel Coelho, Michael Moshe, and Andrew Murray for useful discussions related to this work. AA thanks the A.P. Sloan foundation, the Milton Fund, the Volkswagen Foundation and Harvard Dean's Competitive Fund for Promising Scholarship for their support. JL was supported by the George F. Carrier fellowship and the National Science Foundation through the Harvard Materials Research Science and Engineering Center (DMR-1420570). JM was funded by Monticello Foundation Internship and the Robert and Delpha Noland Summer Internships.

\bibliography{lin}

\clearpage

\section{Supplementary Information}
\renewcommand{\theequation}{A.\arabic{equation}}
\setcounter{equation}{0}

\renewcommand{\thefigure}{A.\arabic{figure}}
\setcounter{figure}{0}

\begin{figure}[htb!]
	\includegraphics[width=.45\textwidth]{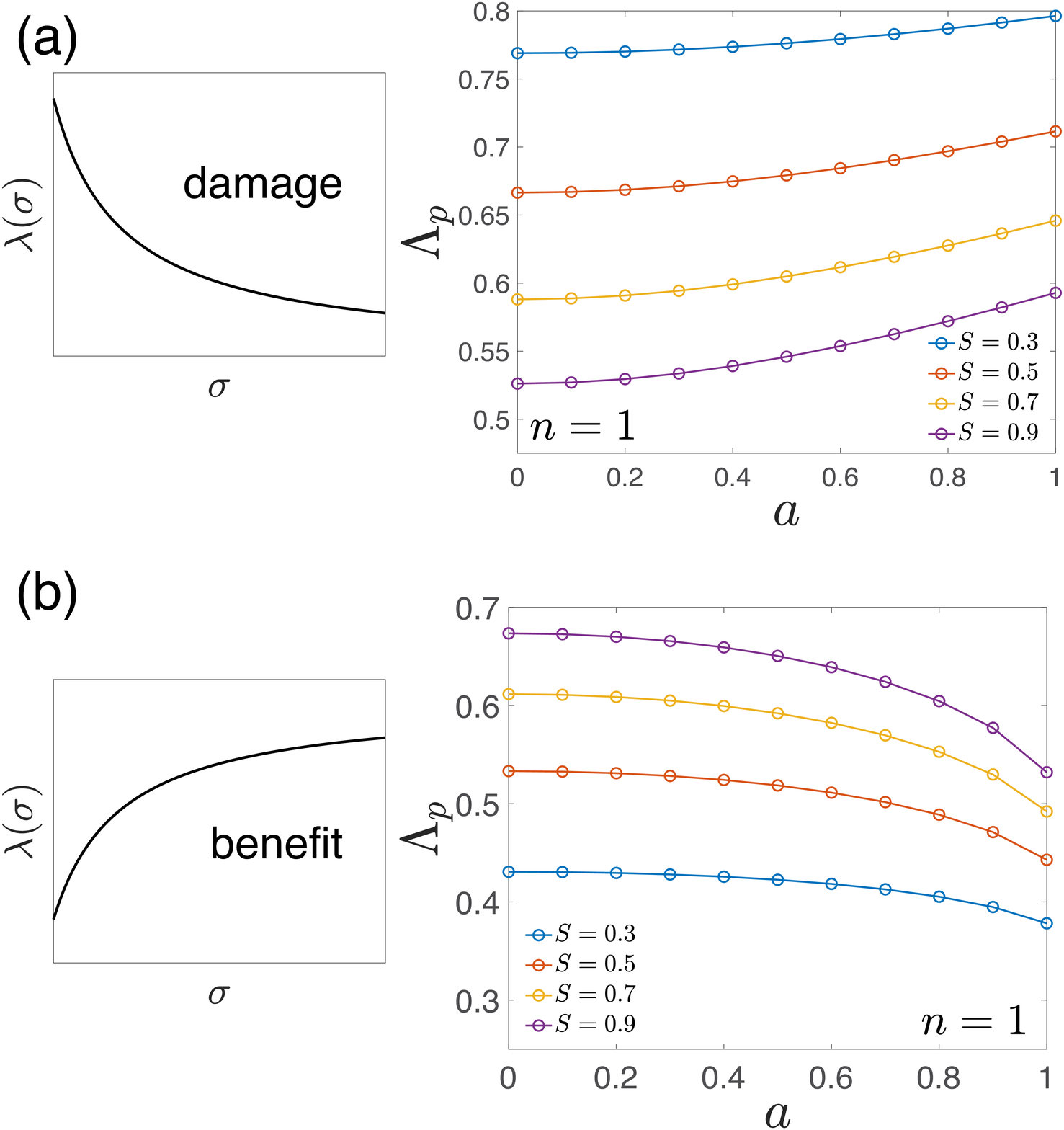}
	\caption{The population growth rate for the $n=1$ case {\it v.s.} the asymmetry parameter $a$ for different protein accumulation rates $S$ (environmental stress levels). For (a), $\lambda[\sigma]=1/(1+\sigma)$ (left). For (b), $\lambda[\sigma]=\sigma/(1+\sigma)+0.2$ (right).}\label{figures1}
\end{figure}

\begin{figure}[htb!]
	\includegraphics[width=.45\textwidth]{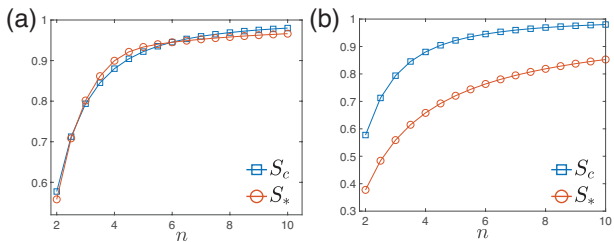}
	\caption{ $S_c$ is the inflection point of the growth rate function. (a) For the damage case, in general $S_c\approx S_{\ast}$ for the whole range of the Hill exponent $n$.  $\lambda_0=1$, $\lambda+1=0$ (d) For the benefit case, $S_{\ast}$ is below $S_c$ for all $n$. $\lambda_0=0.2$, $\lambda_1=1.2$. }\label{figures8}
\end{figure}

\subsection{(a) $n=1$ case}
In the main text, we argue that when the Hill exponent $n=1$, the optimal asymmetry value is $a_c=1$ for the damage case and $a_c=0$ for the benefit case. In Fig. \ref{figures1} we provide numerical evidence supporting this result.

\subsection{(b) $n>2$ cases}
In this section, we show the numerical computations of $S_{\ast}$ for Hill-type growth rate function with Hill exponent $n>2$, see Fig. \ref{figures8}. We find in general $S_{\ast}\approx S_{c}$ for the damage case and $S_{\ast}<S_c$ for the benefit case. Therefore, the conclusions on the nature of phase transition of protein segregation we made in the main text remain valid for $n>2$.
\subsection{(c) Relation between $S_{\ast}$ and $S_c$}
In the main text, we discuss a growth-rate dependence of protein concentration following a generic Hill-function form:
\begin{equation}
	\lambda[\sigma]=\frac{\lambda_0 + \lambda_1 \sigma^n}{1+\sigma^n}.\label{lambda}
\end{equation}

It is important to keep $\lambda_0>0$ for the case of benefit segregation, to avoid vanishing growth rates for $a=1$. However, the precise value of $\lambda_0$ hardly affects the results (as we show in the next section), and in particular, a well defined limit exists when $\lambda_0 \to 0$. In this section, for simplicity and concreteness, we analytically study two particular cases: the dependence $\lambda[\sigma]=1/(1+\sigma^n)$ for the damage case (corresponding to $\lambda_0=1$ and $\lambda_1=0$) and the dependence  $\lambda[\sigma]=\sigma^n/(1+\sigma^n)$ for the benefit case (corresponding to $\lambda_0=0$ and $\lambda_1=1$). 

\begin{figure*}[htb!]
	\includegraphics[width=.95\textwidth]{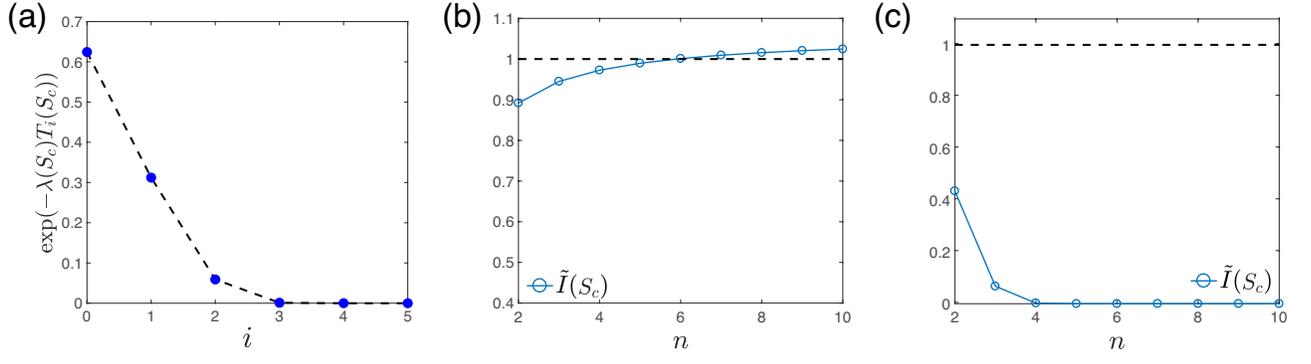}
	\caption{(a) The magnitudes of the first five terms of the sum in Eq. (\ref{sast}) with $S=S_c$ and the Hill exponent $n=3$. (b) The approximate values of $I(S_c)$ ($\tilde{I}(S_c)$), Eq. (\ref{IsI}), {\it v.s.} the Hill exponent $n$. (c) The upper bound of $I(S_c)$ from Eq. (\ref{upper}) for the benefit case.}\label{figures2}
\end{figure*}

We define
\begin{equation}
	I(S)=\sum_{i=0}^{\infty} e^{-\lambda[S] T_i(S)},\label{sast}
\end{equation}
where $T_i(S)=\sum_{j=0}^{i} \tau_j(S)$, and
\begin{equation}
	\tau_j(S)=\int_{1}^{2}dV\frac{dV}{V\lambda[S(1+\frac{j-1}{V})]}.\label{tauj}
\end{equation}
$S_{\ast}$ is the accumulation rate of protein at which $\Lambda_p(S_{\ast},a=0)=\Lambda_p(S_{\ast},a=1)$, therefore $I(S_{\ast})=1$.

We first consider the damage case and show that $I(S_c)\approx 1$ for a general monotonically decreasing function $\lambda[\sigma]$ that has an inflection point. The terms in the sum of Eq. (\ref{sast}) decay quickly since the generation time increases as the damage level at cell birth increases. As we show in Fig. \ref{figures2}(a) for $n=3$, the third term ($i=2$) is already much smaller than the first two terms, so we approximate the sum as
\begin{equation}
	I(S_c)\approx \sum_{i=0}^{1} \exp(-\lambda[S_c] T_i(S_c)) .\label{sI}
\end{equation}

Eq. (\ref{tauj}) implies $\tau_1(s)=\ln(2)/\lambda[S]$, so we can further simplify the above equation to
\begin{equation}
	I(S_c)\approx   \frac{3}{2} e^{-\lambda[S_c]\tau_0(S_c)}.\label{Is}
\end{equation}
The problem reduces to the calculation of $\tau_0(S_c)$. According to Eq. (\ref{tauj}), the protein concentration $\sigma$ involved in the integration is from $0$ to $S_c/2$. Because $S_c$ is the inflection point at which the growth rate starts to drop rapidly, we can approximate the growth rate in the range from $0$ to $S_c/2$ as $1$ and $\tau_0(S_c)\approx \ln(2)$. Finally, Eq. (\ref{Is}) reduces to
\begin{equation}
	I(S_c)\approx \frac{3}{2}2^{-\lambda[S_c]}.\label{IsI}
\end{equation}
So far, we have not assumed any special form of $\lambda[\sigma]$. Now we consider Eq. (\ref{IsI}) with $\lambda[\sigma]=1/(1+\sigma^n)$, with $n\geq 2$. We define $\tilde{I}(S_c)\equiv \frac{3}{2}2^{-\lambda[S_c]}$. Plugging in the inflection point $S_c(n)=(\frac{n-1}{n+1})^{1/n}$ in $\tilde I(S_c)$, we find $\tilde{I}(S_c)\approx 1$ (Fig \ref{figures2}b). This implies that for the damage case, $S_c$ is typically close to $S_{\ast}$, and the transition from $a_c=0$ to $a_c=1$ must be sharp.

Next we consider the benefit case. To prove that $a_c$ undergoes a second order transition, we need to verify that $\Lambda_p(S_c,a=0)>\Lambda_p(S_c,a=1)$, which is equivalent to showing
\begin{equation}
	I(S_c)=\sum_{i=0}^{\infty} e^{-\lambda[S_c] T_i(S_c)}<1.\label{si}
\end{equation}
We rewrite the above equation as
\begin{align}
	I(S_c)&= e^{-\lambda[S_c]\tau_0(S_c)}\nonumber \\
	& \left(\frac{3}{2} + \frac{1}{2} e^{-\lambda[S_c]\tau_2(S_c)}+  \frac{1}{2} e^{-\lambda[S_c](\tau_2(S_c)+\tau_3(S_c))}+ ...\right)
\end{align}
Note that we have not made any approximation so far. The upper bound of the growth rate is $1$, and the lower bound of generation time is $\ln(2)$ as the initial benefit level increases. Using the lower bound of generation time, we can find an upper bound for $I(S_c)$,
\begin{align}
	I(S_c)&< e^{-\lambda[S_c]\tau_0(S_c)} \left(1+\frac{1}{2} + \frac{1}{2}\sum_{n=1}^{\infty} 2^{-n\lambda[S_c]}\right)\nonumber \\
	&= e^{-\lambda[S_c]\tau_0(S_c)} \left(1+\frac{1}{2(1-2^{-\lambda[S_c]})}\right).
\end{align}
Because $\tau_0(S_c)>\ln(2)/\lambda[S_c/2]$, we find,
\begin{equation}
	I(S_c)<2^{-\frac{\lambda[S_c]}{\lambda[S_c/2]}} \left(1+\frac{1}{2\left(1-2^{-\lambda[S_c]}\right)}\right).\label{upper}
\end{equation}
The upper bound, denoted as $\tilde{I}(S_c)$, is indeed smaller than 1 (Fig. \ref{figures2}c). Therefore, $\Lambda_p(S_c,a=0)>\Lambda_p(S_c,a=1)$, and the transition must be second order.

\begin{figure}[htb!]
	\includegraphics[width=.48\textwidth]{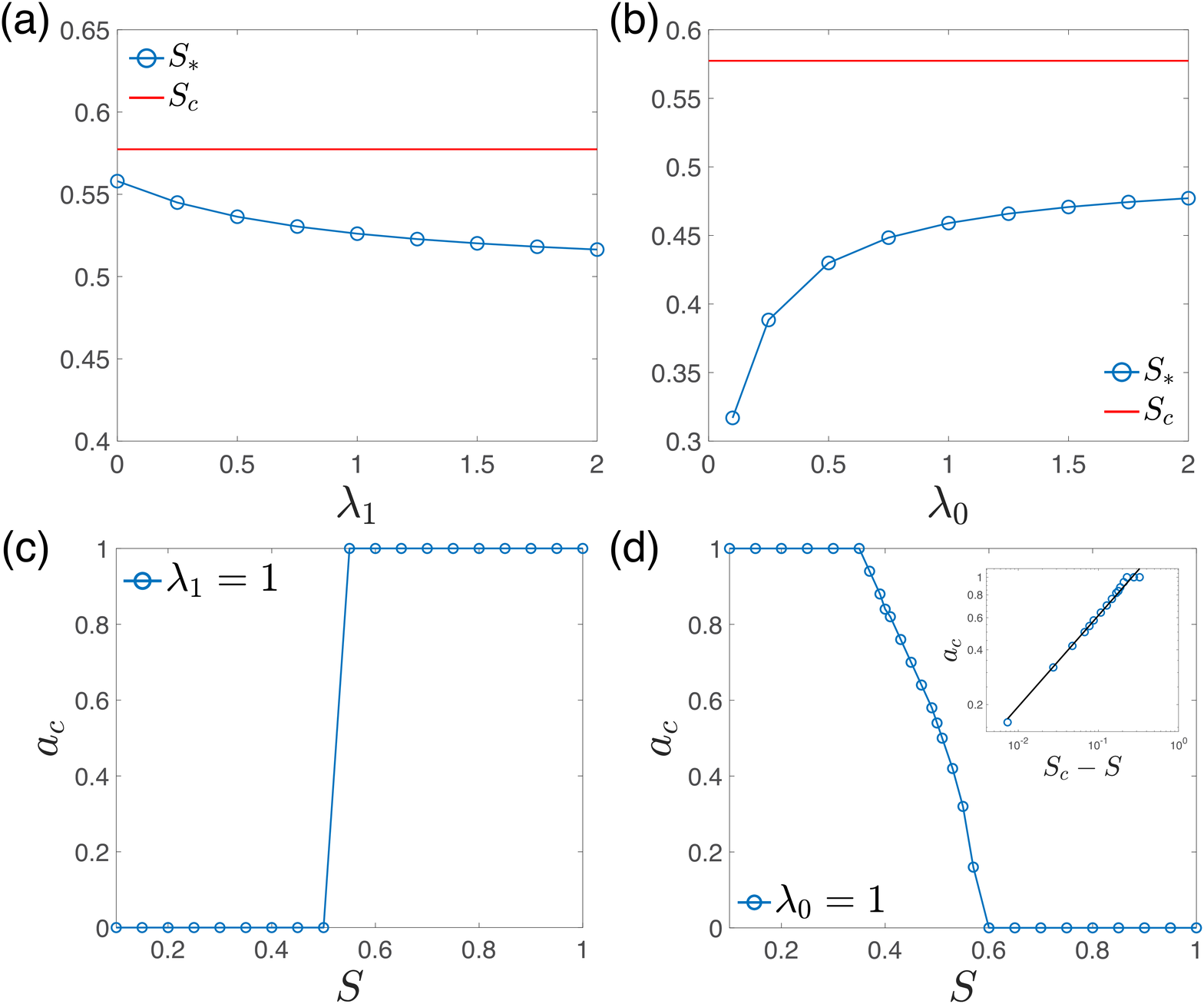}
	\caption{(a) $S_{\ast}$ {\it v.s.} $\lambda_1$. The growth rate function is $\lambda[\sigma]=\frac{1}{1+\sigma^2}+\lambda_1$. (b)  $S_{\ast}$ {\it v.s.} $\lambda_0$. The growth rate function is $\lambda[\sigma]=\frac{\sigma^2}{1+\sigma^2}+\lambda_0$. (c, d) The optimal asymmetry degree $a_c$ {\it v.s.} the accumulation rate $S$ for $\lambda_1=1$ (c, damage) and $\lambda_0=1$ (d, benefit). The inset shows the mean field scaling near the critical point ($S_c\approx 0.577$) where the black line has a slope $1/2$.}\label{figures3}
\end{figure}

\subsection{(d) Effects of residual growth rate in the Hill type growth rate function}
In the main text, we mainly considered the situation when $\lambda[\sigma]=\frac{1}{1+\sigma^n}$ for the damage segregation case and $\lambda[\sigma]=\frac{\sigma^n}{1+\sigma^n}+0.2$ for the benefit segregation case. These choices correspond to $\lambda_0=1, \lambda_1=0$ for the damage case and $\lambda_0=0.2, \lambda_1=1.2$ for the benefit cases according to $\lambda[\sigma]=\frac{\lambda_0+\lambda_1\sigma^2}{1+\sigma^2}$ (Eq. (4) of the main text). Here we generalize the growth rate function by considering other choices of $\lambda_1$  for the damage case and $\lambda_0$ for the benefit case. To simplify the notation, we take the growth rate function as $\lambda[\sigma]=\frac{1}{1+\sigma^n}+\lambda_1$ for the damage case and $\lambda[\sigma]=\frac{\sigma^n}{1+\sigma^n}+\lambda_0$. We numerically compute $S_{\ast}$ based on Eq. (5) in the main text, and compare it with the inflection point $S_c$. As shown in Fig. \ref{figures3}, $S_{\ast}$ decreases as $\lambda_1$ increases, and is always smaller than $S_c$ for a wide range of $\lambda_1$ for the damage segregation. For the benefit case, $S_{\ast}$ increases as $\lambda_0$ increases, but always smaller than $S_c$. In conclusion, we find that the sharp transition nature of damage segregation and smooth transition nature of benefit segregation is intact as the residual growth rate $\lambda_1$ or $\lambda_0$ changes. To confirm this, we also show the optimal asymmetry degree of segregation $a_c$ in Fig. \ref{figures3}(c,d), which is indeed quite sharp for the damage segregation and continuous for the benefit segregation.

\subsection{(e) Fitting coefficients of fitness function}
\begin{figure}[htb!]
	\includegraphics[width=.45\textwidth]{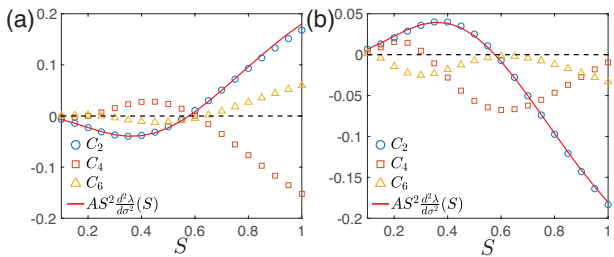}
	\caption{The fitting coefficients of the fitness as function of $a$ using $f=C_2a^2+C_4a^4+C_6a^6$ for the damage (a) and benefit (b) cases from Fig. 3(c,d) in the main text. The red lines are the theoretical predictions from the Landau approach, and $A\approx 0.361$ for both cases.}\label{figures7}
\end{figure}

In this section, we show the fitting coefficients, $C_2$, $C_4$, and $C_6$ of the fitness function based on the Landau approach, $f=C_2a^2+C_4a^4+C_6a_6$ (Fig. \ref{figures7}). We find that in both cases, a universal $A=0.361$ can match the theoretical prediction Eq. (6) in the main text ($C_2=AS^2d^2\lambda/d\sigma^2(S)$) well. In the next section, we will show the value of $A$ is indeed universal and independent of the specific growth rate function.

\begin{figure}[htb!]
	\includegraphics[width=.5\textwidth]{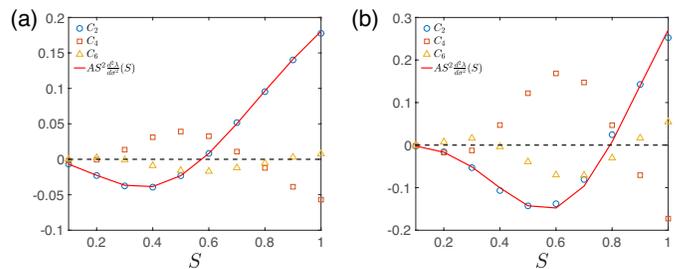}
	\caption{The fitting coefficients of the idealized population. Here $A=0.361$ for both $\lambda[\sigma]=1/(1+\sigma^2)$ (a) and $\lambda[\sigma]=1/(1+\sigma^3)$ (b), the same as the numerical value shown in Fig. \ref{figures7}(a,b).}\label{figures4}
\end{figure}
\subsection{(f) Universality of the lowest order coefficient in the Landau approach}
For any $a>0$, the dependence of growth rate on the protein concentration perturbs the structure of the population tree: since cellular doubling times are affected, the cell division timing is now different even across cells of the same generation. In this section, we will consider an approximation whereby we \emph{neglect} this effect, and show that we can accurately compute the lowest order coefficient $A$ in the Landau expansion (Eq.(6) in the main text), which we show to be independent of the choice of growth rate function.

Within our approximation scheme, we consider an idealized population in which the key protein is segregated with the asymmetry parameter $a$, while it does not affect the growth rate (i.e., all cells grow at a constant rate $\lambda=1$). We will find the protein concentration of each cell for this idealized scenario, and compute the population growth rate $\Lambda_p^{\prime}$ taking into account the growth rate dependence on the protein concentration.

\begin{equation}
	\Lambda_p^{\prime}=\frac{\sum_i \lambda[\sigma_i]V_i}{\sum V_i}.
\end{equation}
We take into account the age distribution of cells for an asynchronous population with a fixed generation time, $P(\tau)=\ln(2)2^{1-\tau}$, where $\tau$ is the normalized age from $0$ to $1$ \cite{Powell1956}. Therefore the corresponding cell volume distribution is $P(V)=2/V^2$ with $1\leq V\leq 2$. We take two different growth rate functions, $\lambda[\sigma]=1/(1+\sigma^2)$, $\lambda[\sigma]1/(1+\sigma^3)$, and numerically fit $f=\Lambda_p^{\prime}(S,a)-\lambda[S]$ as a function of $a$ using $f=C_2a^2+C_4a^4+C_6a^6$. We find that even though the population tree structure is completely deterministic (within our approach) and independent of the growth rate function, the resulting population growth rate $\Lambda_p^{\prime}$ shows the same asymptotic form in the small $a$ limit, $f\rightarrow A S^2\frac{d^2\lambda}{d\sigma^2}a^2$, with $A\approx 0.361$ for both of the two growth rate functions (Fig. \ref{figures4}). This evidence supports that in the limit of small $a$, the dominant effects of $a$ on the population growth rate is through the growth rate function $\lambda[\sigma]$ rather than changing the population tree structure.
\begin{figure}[htb!]
	\includegraphics[width=.45\textwidth]{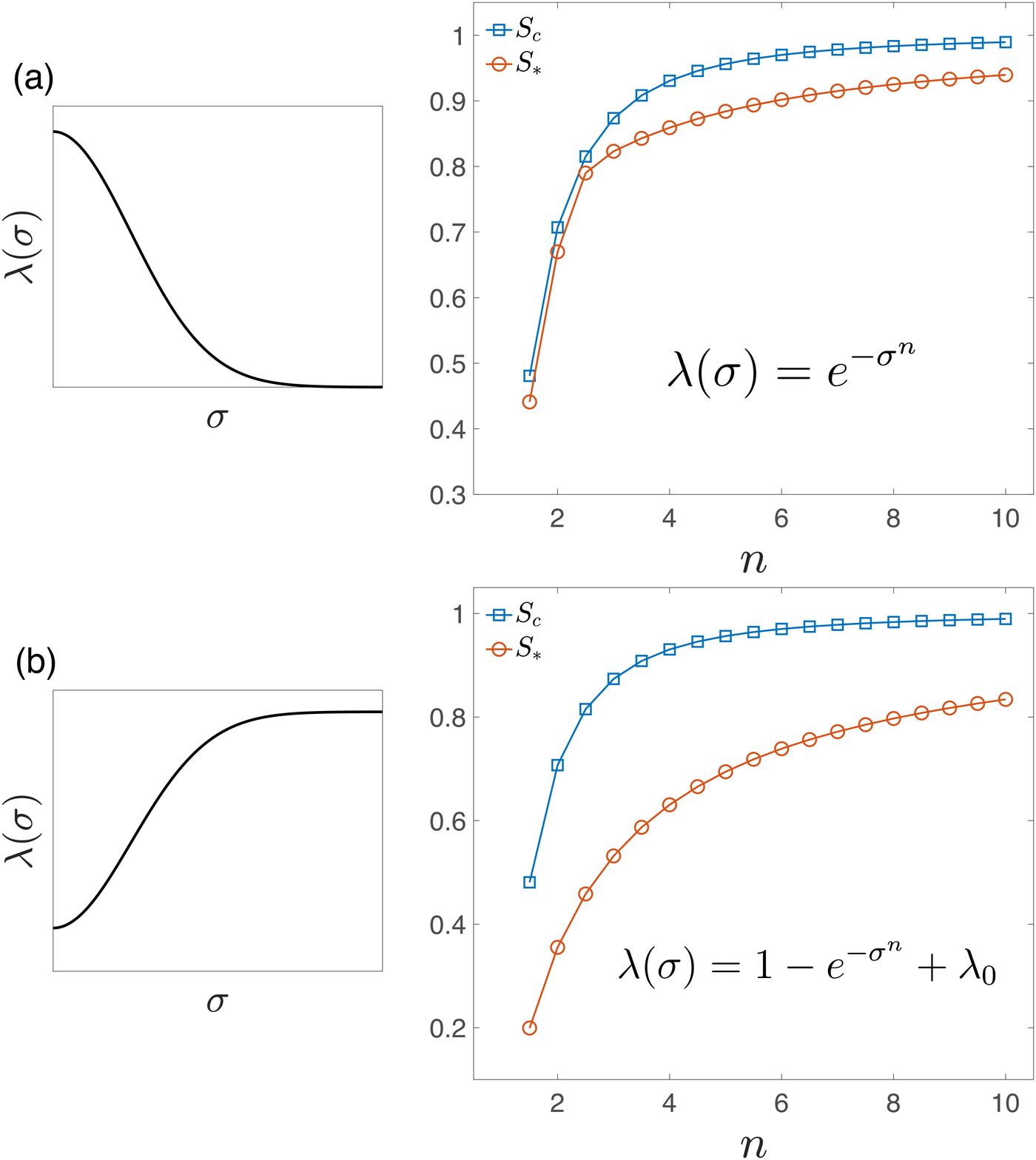}
	\caption{The growth rate functions are shown in the figures and examples with $n=2$ are shown on the left. $S_c$ is the inflection point of the growth rate function at which its second derivative vanishes. (a) For the damage case, in general $S_{\ast}\lesssim S_{c}$ for the whole range of the exponent $n$. (b) For the benefit case, $S_{\ast}$ is far below $S_c$ for all $n$. $\lambda_0=0.2$.}\label{figures5}
\end{figure}

\subsection{(g) Other growth rate functions}
In the main text, we mainly consider the Hill type growth rate function and use the Landau approach to study the transition from the symmetric phase to asymmetric phase. The main conclusion is that for the damage case, if the two special accumulation rates satisfy $S_{\ast} < S_c$ (or $S_{\ast}\approx S_c$), the transition from $a_c=0$ to $a_c=1$ is sharp. For the benefit case, if $S_{\ast}<S_c$, $a_c$ undergoes a mean field second order transition as $S$ changes. These arguments are independent of the growth rate function. Here we show that the relations between $S_{\ast}$ and $S_c$ for another type of growth rate function (Fig. \ref{figures5})
\begin{subequations}
	\begin{align}
		\lambda[\sigma]&=\exp(-\sigma^n),\\
		\lambda[\sigma]&=1-\exp(-\sigma^n)+\lambda_0,
	\end{align}\label{lambda}
\end{subequations}
also satisfy these conditions.

\end{document}